\documentclass[fleqn,12pt]{article}
\usepackage{amsmath,amssymb,epsfig,times,graphics}
\begin{document}

\title{\bf A multi-phenotypic cancer model with cell plasticity}

\date{}
\maketitle

\author{Da Zhou$^{1,*}$, Yue Wang$^2$, Bin Wu$^3$}

\begin{enumerate}
  \item School of Mathematical Sciences, Xiamen University,
Xiamen 361005, P.R. China\\ ($*$Corresponding Author, zhouda@xmu.edu.cn)
  \item Department of Applied Mathematics, University of Washington,
Seattle, WA 98195, USA
  \item Evolutionary Theory Group, Max-Planck-Institute for Evolutionary Biology,
August-Thienemann-Stra$\beta$e 2, 24306  Pl{\"o}n, Germany
\end{enumerate}

\begin{abstract}

The conventional cancer stem cell (CSC) theory indicates a hierarchy of CSCs and non-stem cancer cells (NSCCs),
that is, CSCs can differentiate into NSCCs but not vice versa. However, an alternative paradigm of CSC theory
with reversible cell plasticity among cancer cells has received much attention very recently. Here we present a
generalized multi-phenotypic cancer model by integrating cell plasticity with the conventional hierarchical structure
of cancer cells. We prove that under very weak assumption, the nonlinear dynamics of multi-phenotypic proportions
in our model has only one stable steady state and no stable limit cycle. This result theoretically explains the
phenotypic equilibrium phenomena reported in various cancer cell lines. Furthermore, according to the
transient analysis of our model, it is found that cancer cell plasticity plays an essential role in
maintaining the phenotypic diversity in cancer especially during the transient dynamics.
Two biological examples with experimental data show that
the phenotypic conversions from NCSSs to CSCs greatly contribute to the transient growth of CSCs proportion shortly
after the drastic reduction of it. In particular, an interesting overshooting phenomenon of CSCs proportion arises in
three-phenotypic example. Our work may pave the way for modeling and analyzing the multi-phenotypic
cell population dynamics with cell plasticity.

\end{abstract}

\newpage

%
%


\section{Introduction}
\label{}

The cancer stem cell (CSC) hypothesis \cite{reya2001stem,jordan2006cancer}
states that tumors or hematological cancers arise from a small number of
stem-like cancer cells with the abilities of self-renewal and differentiation into
other non-stem cancer cells (NSCCs).
That is, the conventional cancer stem cell theory suggests a cellular
hierarchy where CSCs are at the apex \cite{dalerba2007cancer}. Based on this paradigm,
cancer stem cell models were widely investigated in previous literature on theoretical biology
\cite{colijn2005mathematical,johnston2007mathematical,boman2007symmetric,dingli2007symmetric,johnston2010proportion,
antal2011exact,sottoriva2011modeling,werner2011dynamics,stiehl2012mathematical,molina2012simple}.

However, recent studies have highlighted the complexities and challenges in the evolving concept of CSC
\cite{nguyen2012cancer, visvader2012cancer}. In particular, it was reported that reversible phenotypic
changes can occur between stem-like cancer cells and more differentiated cancer cells.
Meyer \emph{et al} showed that interconversion occurred both \emph{in vivo} and \emph{in vitro} between
noninvasive, epithelial-like $CD44^+CD24^+$ cells and invasive, mesenchymal $CD44^+CD24^-$
cells in breast cancer \cite{meyer2009dynamic}.
Chaffer \emph{et al} showed both \emph{in vivo} and \emph{in vitro}
that transformed (oncogenic) $CD44^{lo}$-HMECs (human mammary epithelial cells)
could spontaneously convert to $CD44^{hi}$-CSCs \cite{chaffer2011normal}.
The results by Quintana \emph{et al} indicated that phenotypically diverse cancer cells
in both primary and metastatic melanomas
can undergo reversible phenotypic conversions \emph{in vivo} \cite{quintana2010phenotypic}.
Scaffidi \emph{et al} showed that
stem-like cancer cells can be generated \emph{in vitro}
from transformed (oncogenic) fibroblasts during neoplastic transformation \cite{scaffidi2011vitro}.
The conversion from NSCCs to CSCs was also \emph{in situ} visualized by Yang \emph{et al} in SW620 colon cancer cell line
\cite{yang2012dynamic}.
Besides, the phenotypic transitions between different NSCCs
in breast cancer was reported both \emph{in vivo} and \emph{in vitro}
\cite{gupta2011stochastic}.
These various types of phenotypic conversions among cancer cells, also known as
\emph{cancer cell plasticity} mechanisms \cite{french2012complex}, provide new thinking about the CSC hypothesis
and related therapeutic strategy \cite{pisco2013non}.

Special attention has recently been paid to the mathematical models concerning cancer cell plasticity.
Gupta \emph{et al} introduced a simple Markov chain model of stochastic transitions among
stem-like, basal and luminal cells in breast cancer cell lines \cite{gupta2011stochastic}.
Zapperi and La Porta compared mathematical models for cancer cell
proliferation that take phenotypic switching and imperfect biomarker into account
\cite{zapperi2012cancer}. A series of works by
dos Santos and da Silva developed a model with the effects of stochastic noise and cell plasticity
for explaining the variable frequencies of CSCs in tumors \cite{dos2013possible,dos2013noise}.
Zhou \emph{et al} compared the transient dynamics of the bidirectional and unidirectional models
of CSCs and NSCCs \cite{zhoupopulation}.
Wang \emph{et al} showed that tumor heterogeneity may exist in the model
with both CSC hierarchy and cell plasticity \cite{wang2014dynamics}.
Leder \emph{et al}'s model described the reversible phenotypic interconversions between
the stem-like resistant cells (SLRCs) and the differentiated
sensitive cells (DSCs) in glioblastomas, which revealed
optimized radiation dosing schedules \cite{leder2014mathematical}. These works demonstrated that
cell plasticity provides new insight into cancer cell population dynamics.

To further explore how cell plasticity challenges the hierarchical cancer stem cell scenario,
and in particular how cell plasticity influences tumor heterogeneity,
one should incorporate cell plasticity into the development of cancer which
is full of biological complexities. One particular and crucial complexity
arises from highly diverse phenotypes in the population of cancer cells. The aforementioned
mathematical models were mainly focused on the relation between CSCs and NSCCs
\cite{zapperi2012cancer,dos2013possible,dos2013noise,zhoupopulation,wang2014dynamics,leder2014mathematical},
that is, their models simply classified the cancer cells into two ``opposite'' phenotypes. This simplification
is effective for studying the reversible conversions between NSCCs and CSCs, but covers up various
phenotype switchings between different cancer cells that are worthy of studying.
As an exceptional case, Gupta \emph{et al}'s model \cite{gupta2011stochastic} consists of three phenotypes
(stem-like, basal and luminal cells), but rigorous mathematical analysis for general multi-phenotypic models
with cell plasticity is still lacking.

In this study, we try to provide a multi-phenotypic framework for integrating cell plasticity with conventional
growth model of cancer cells. A generalized model comprising $1+m$ cellular phenotypes
(one CSC phenotype and $m$ different NSCC phenotypes) is put forward.
Besides cell-state conversions from NSCCs to CSCs and phenotype switchings
between different NSCC phenotypes,
the cellular processes in classical cancer stem cell models
(\emph{i.e.}, asymmetric cell division, symmetric cell division and cell death)
are also included in our model.
When the cell population size is not large enough and subject to stochastic fluctuations, our model is formulated by
a continuous-time high-dimensional Markov process. In the limit of large population size, the model can be
governed by a system of linear ordinary differential equations (ODEs).
Moreover, to investigate the dynamics of phenotypic proportions, the population model is converted
into a nonlinear frequency one. It is shown that under very weak assumption,
the nonlinear frequency model has only one stable steady state and no stable limit cycle.
Not only does this result theoretically explain the \emph{phenotypic equilibrium} phenomena
reported in various cancer cell lines
\cite{chaffer2011normal,yang2012dynamic,gupta2011stochastic,iliopoulos2011inducible},
but it is also predicted that the phenotypic equilibrium should be universal in the population of
multi-phenotypic cancer cells.
Furthermore, it is also found that cancer cell plasticity
greatly influences the transient proportions of cell phenotypes. In particular,
two concrete examples with experimental data are presented, showing that
the cell-state conversions play an essential role in the transient growth of CSCs proportion
shortly after the drastic reduction of it. In particular,
an interesting \emph{overshooting} phenomenon of CSCs proportion arises in S-B-L model with cell plasticity.
Note that two-phenotypic models never perform overshooting \cite{zhoupopulation, jia2013overshoot},
overshooting can be a result of interplay between cell plasticity and diversity of phenotype. Moreover,
it has been investigated in ecology and population genetics that
phenotypic variability can serve as an advantageous strategy for biological populations in fluctuating environments \cite{wolf2005diversity,kussell2005phenotypic,lu2007phenotypic,acar2008stochastic,kaneko2012evolution},
our findings thus enrich this idea that cell plasticity as a surviving strategy might be more essential in
maintaining the phenotype diversity (heterogeneity) of cancer especially during transient dynamics.

The paper is organized as follows. The model framework is formulated in Section 2.
In Section 3, we investigate the frequency model and phenotypic equilibrium. The roles
of cell-state conversions in transient dynamics are discussed in Section 4.
Conclusions are presented in Section 5.

\section{Model description}
\label{}

\subsection{Assumptions}

This section describes the assumptions of the model investigated in this study.
Consider a population of cancer cells comprising $1+m$ phenotypes:
$\textrm{CSC}$ represents cancer stem cell, and
$\textrm{NSCC}_1$, $\textrm{NSCC}_2$, ..., $\textrm{NSCC}_m$ represent $m$
different phenotypes of non-stem cancer cells. In this model,
cell plasticity is integrated with the growth model of cancer cells.
According to conventional cancer stem cell scenario,
not only can CSC divide asymmetrically into
two unequal daughter cells (one CSC and one NSCC) \cite{reya2001stem}, but it can
also divide symmetrically into two daughter CSCs \cite{todaro2010colon}
\footnote{It should be noted that, another type of symmetric division
that CSC divides into two daughter NSCCs (termed symmetric differentiation \cite{morrison2006asymmetric})
is not accounted for in our model,
however it is shown in \ref{appendixE} that the main results
achieved in this study are still valid for the model including symmetric differentiation,
implying the kinetic equivalence between the two models.}.
So for CSC we assume that
\begin{itemize}
  \item \textbf{Symmetric division:} CSC $\overset{\alpha_{00}}{\longrightarrow}$ CSC+CSC;
  \item \textbf{Asymmetric division:} CSC $\overset{\alpha_{0j}}{\longrightarrow}$ CSC+$\textrm{NSCC}_j$ ($1\leq j \leq m$);
  \item \textbf{Cell death:} CSC $\overset{\alpha_{0}}{\longrightarrow}$ $\emptyset$.
\end{itemize}
For NSCCs, besides the symmetric division, two types of
cell plasticity mechanisms that have been reported in previous biological literature
are included in our model, \emph{i.e.}, the cell-state conversions from
NSCCs to CSCs (termed\emph{ de-differentiation}) \cite{yang2012dynamic} and phenotype switchings between different NCSSs
\cite{gupta2011stochastic}. In this way, for $\textrm{NSCC}_i$ ($1\leq i \leq m$) we assume that
\begin{itemize}
  \item \textbf{Symmetric division:} $\textrm{NSCC}_i$ $\overset{\alpha_{ii}}{\longrightarrow}$ $\textrm{NSCC}_i$+$\textrm{NSCC}_i$;
  \item \textbf{De-differentiation:} $\textrm{NSCC}_i$ $\overset{\alpha_{i0}}{\longrightarrow}$ CSC;
  \item \textbf{phenotype switching:} $\textrm{NSCC}_i$ $\overset{\alpha_{ij}}{\longrightarrow}$ $\textrm{NSCC}_j$ ($i\neq j$);
  \item \textbf{Cell death:} $\textrm{NSCC}_i$ $\overset{\alpha_{i}}{\longrightarrow}$ $\emptyset$.
\end{itemize}
We list the elements of the model in table \ref{table1} (see Fig. 1 for the example of three phenotypes).
If we denote $X^0_t$, $X^1_t$,..., $X^m_t$ as the cell numbers of $\textrm{CSC}$, $\textrm{NSCC}_1$, ..., $\textrm{NSCC}_m$ at time $t$
respectively, then $\overrightarrow{X_t}=(X^0_t, X^1_t,..., X^m_t)^T$ is an ($m+1$)-dimensional cellular system consisting of
$(m+1)\times(m+2)$ cellular processes.
On the basis of the modeling principles in biochemical reactions (see Chapter 7 in \cite{van1992stochastic}),
our cellular system can be formulated by both stochastic and deterministic models as follows:

\begin{table}[htbp]\small
\caption{Parameters used in the model}
\begin{tabular}{|c|c|c|}
  \hline
  \hline
  \textbf{Symbol} & \textbf{Parameter} & \textbf{Cellular process}\\
  \hline
  $\alpha_{00}$ & symmetric division rate by CSC & CSC $\overset{\alpha_{00}}{\longrightarrow}$ CSC+CSC\\
  \hline
  $\alpha_{0j}$ & asymmetric division rate by CSC & CSC $\overset{\alpha_{0j}}{\longrightarrow}$ CSC+$\textrm{NSCC}_j$ ($1\leq j \leq m$)\\
  \hline
  $\alpha_{0}$ & death rate by CSC & CSC $\overset{\alpha_{0}}{\longrightarrow}$ $\emptyset$\\
  \hline
  $\alpha_{ii}$ & symmetric division rate by $\textrm{NSCC}_i$ & $\textrm{NSCC}_i$ $\overset{\alpha_{ii}}{\longrightarrow}$ $\textrm{NSCC}_i$+$\textrm{NSCC}_i$ ($1\leq i \leq m$)\\
  \hline
  $\alpha_{i0}$ & phenotypic conversion rate from $\textrm{NSCC}_i$ to CSC & $\textrm{NSCC}_i$ $\overset{\alpha_{i0}}{\longrightarrow}$ CSC ($1\leq i \leq m$)\\
  \hline
  $\alpha_{ij}$ & phenotypic conversion rate from $\textrm{NSCC}_i$ to $\textrm{NSCC}_j$ & $\textrm{NSCC}_i$ $\overset{\alpha_{ij}}{\longrightarrow}$ $\textrm{NSCC}_j$ ($i\neq j$)\\
  \hline
  $\alpha_{i}$ & death rate by $\textrm{NSCC}_i$ & $\textrm{NSCC}_i$ $\overset{\alpha_{i}}{\longrightarrow}$ $\emptyset$ ($1\leq i \leq m$)\\
  \hline
  \hline
\end{tabular}
\label{table1}
\end{table}

\subsection{Stochastic model}

The randomness of cellular systems is essentially rooted in the stochastic nature
of gene expression in individual cells \cite{cai2006stochastic}.
The emergence of determinism from randomness is a result of the \emph{Law of Large Numbers}. In other words, when
the population is not large enough and subject to stochastic fluctuations,
$\overrightarrow{X}=(X^0_t, X^1_t,..., X^m_t)^T$ should be stochastic.
According to the theory of \emph{Chemical Master Equation} (CME) (see Chapter 11 in \cite{beard2008chemical}),
the stochastic model can be formulated by
a continuous-time Markov process on $\mathbb{N}^{m+1}$.
If we define $\textrm{Pr}\{x_0,...,x_i,...,x_m; t\}$ as the probability of $\overrightarrow{X}=(x_0,...,x_i,...,x_m)^T$ at time $t$,
then the rate of change in $\textrm{Pr}\{x_0,...,x_i,...,x_m; t\}$ is equal to the rate of transition from all the other possible states to
$(x_0,...,x_i,...,x_m)$ minus the rate of transition from $(x_0,...,x_i,...,x_m)$ to all the other possible states, formally,
\begin{eqnarray}
\frac{d \textrm{Pr}\{x_0,...,x_i,...,x_m; t\}}{d t}=F(x_0,...,x_i,...,x_m, t),
\label{Masterequation}
\end{eqnarray}
where
\begin{eqnarray*}
F(x_0,...,x_i,...,x_m, t)&=&\sum_{i=0}^m(x_i-1)\alpha_{ii}\textrm{Pr}\{x_0,...,x_i-1,...,x_m; t\}\\
&&+\sum_{i=0}^m(x_i+1)\alpha_{i}\textrm{Pr}\{x_0,...,x_i+1,...,x_m; t\}\\
&&+\sum_{j=1}^m(x_0\alpha_{0j}\textrm{Pr}\{x_0,...,x_j-1,...,x_m; t\})\\
&&+\sum_{\substack{(i,j)\in \{1,2,...,m\}\times \{0,1,...,m\}\\ \textrm{with}~i\neq j}}(x_i+1)\alpha_{ij}\textrm{Pr}\{x_0,...,x_i+1,...,x_j-1...,x_m; t\}\\
&&-\sum_{(i,j)\in \{0,1,...,m\}\times \{0,1,...,m\}}(x_i\alpha_{ij})\textrm{Pr}\{x_0,...,x_i,...,x_m; t\}\\
&&-\sum_{i=0}^{m}(x_i\alpha_{i})\textrm{Pr}\{x_0,...,x_i,...,x_m; t\}.
\end{eqnarray*}
The first four items of $F(x_0,...,x_i,...,x_m, t)$ describe the transitions from other states to $(x_0,...,x_i,...,x_m)$. For example, the first item
corresponds to the sum of the rates of transition from $(x_0,...,x_i-1,...,x_m)$ to $(x_0,...,x_i,...,x_m)$ by symmetric division
of $i^{th}$ cell phenotype, $0\leq i \leq m$. Similarly, the other three items can be explained by other types of cellular processes accordingly. The last two items of
$F(x_0,...,x_i,...,x_m, t)$ describe the transitions from $(x_0,...,x_i,...,x_m)$ to other states.
Since it is difficult to calculate the analytic solution of $\textrm{Pr}\{x_0,...,x_i,...,x_m; t\}$,
stochastic simulation (called \emph{Gillespie algorithm} \cite{gillespie2007stochastic})
can often be used as an efficient numerical approach to investigate the CME.

When the population size is large enough, it is more convenient to use
deterministic equations to capture the model:

\subsection{Deterministic model}

Let
$\langle\overrightarrow{X_t}\rangle=(\langle X^0_t\rangle, \langle X^1_t\rangle,..., \langle X^m_t\rangle)^T$
be the expectation of $\overrightarrow{X_t}$, where
$$\langle X^i_t\rangle:=\sum_{x_0,...,x_i,...,x_m}x_i\textrm{Pr}\{x_0,...,x_i,...,x_m; t\}~~(0\leq i \leq m).$$
We now derive the equation governing the dynamics of $\langle X^i_t\rangle$
(our method is based on Section 5.8 in \cite{van1992stochastic}).
We multiply $x_i$ on the both sides of
(\ref{Masterequation}), and then calculate the summation over all the indexes
\begin{eqnarray*}
\sum_{x_0,...,x_i,...,x_m}(x_i\frac{d \textrm{Pr}\{x_0,...,x_i,...,x_m; t\}}{d t})=\sum_{x_0,...,x_i,...,x_m}(x_i F(x_0,...,x_i,...,x_m, t)),
\end{eqnarray*}
that is,
\begin{eqnarray*}
\frac{d \langle X^i_t\rangle}{d t}=\sum_{x_0,...,x_i,...,x_m}(x_i F(x_0,...,x_i,...,x_m, t)).
\end{eqnarray*}
Then we have\\
1) when $i=0$\\
\begin{eqnarray}
\frac{d \langle X^0_t\rangle}{d t}=(\alpha_{00}-\alpha_{0})\langle X^0_t\rangle+\sum_{i=1}^m(\alpha_{i0}\langle X^i_t\rangle)).
\end{eqnarray}
2) when $1 \leq i \leq m$\\
\begin{eqnarray}
\frac{d \langle X^i_t\rangle}{d t}=\sum_{j\neq i}\alpha_{ji}\langle X^j_t\rangle+(\alpha_{ii}-\alpha_{i}-\sum_{j\neq i}\alpha_{ij})\langle X^i_t\rangle.
\end{eqnarray}
Thus the dynamics of $\langle\overrightarrow{X_t}\rangle$ can be formulated
by the following $(m+1)$-dimensional linear ordinary differential equations (ODEs)
\begin{eqnarray}
\frac{d \langle\overrightarrow{X_t}\rangle}{d t}=A\langle\overrightarrow{X_t}\rangle,
\label{ODE}
\end{eqnarray}
where
\begin{equation}
A=\left(\begin{smallmatrix}
\alpha_{00}-\alpha_{0} & \alpha_{10} & \alpha_{20} & \cdots & \alpha_{m0}\\
\alpha_{01} & \alpha_{11}-\alpha_{1}-\sum_{j\neq 1}\alpha_{1j} & \alpha_{21} & \cdots & \alpha_{m1} \\
\vdots & \vdots & \vdots & \vdots & \vdots  \\
\alpha_{0(m-1)} & \cdots & \cdots &  \alpha_{(m-1)(m-1)}-\alpha_{(m-1)}-\sum_{j\neq (m-1)}\alpha_{(m-1)j} & \alpha_{m(m-1)} \\
\alpha_{0m} & \cdots & \cdots & \alpha_{(m-1)m} & \alpha_{mm}-\alpha_{m}-\sum_{j\neq m}\alpha_{mj} \\
\end{smallmatrix}\right).
\label{matrix}
\end{equation}
It is noteworthy that Eq. (\ref{ODE}) can also be obtained based on
the \emph{Law of Mass Action} (LMA) \cite{koudrjavcev2001law}, indicating the equivalent kinetic foundation between
the stochastic theory of CME and its deterministic counterpart of LMA.
If we choose to ignore cancer cell plasticity in the model,
\emph{i.e.} let $\alpha_{ij}=0$ $(1\leq i\leq m, 0\leq j\leq m$, and $i\neq j)$, the form of $A$ will reduce to the one governing
the hierarchical cancer stem cell model:
\begin{equation}
A^*=\left(\begin{array}{ccccc}
\alpha_{00}-\alpha_{0} & 0 & 0 & \cdots & 0\\
\alpha_{01} & \alpha_{11}-\alpha_{1} & 0 & \cdots & 0 \\
\vdots & \vdots & \vdots & \vdots & \vdots  \\
\alpha_{0(m-1)} & \cdots & \cdots &  \alpha_{(m-1)(m-1)}-\alpha_{(m-1)} & 0 \\
\alpha_{0m} & \cdots & \cdots & 0 & \alpha_{mm}-\alpha_{m} \\
\end{array}\right).
\label{matrix2}
\end{equation}
In comparison of $A$ and $A^*$, it is interesting that there seems to be a \emph{trade-off} between the diagonal and
off-diagonal elements, that is, $A$ has larger off-diagonal elements but smaller diagonal elements than $A^*$.
Note that the diagonal elements correspond to the growth contributions from the cells with the same phenotypes themselves
while the off-diagonal elements are the contributions from other phenotypes,
cancer cell plasticity can be interpreted as a \emph{altruism behavior} among cancer cells in the sense that
it makes cancer cells of one phenotype help other phenotypes to survive by sacrificing their own
phenotype \cite{axelrod2006evolution}. How this altruism behavior affects the population structure of cancer cells is one
of the major tasks in the study of cell plasticity models.

In the following two sections, we will analyze both the asymptotic and transient dynamics of the model, then apply the theoretical
results to explain the problems arising from concrete biological experiments.

\section{Phenotypic equilibrium and frequency model}

It has been reported in various cancer cell lines
\cite{chaffer2011normal,yang2012dynamic,gupta2011stochastic,iliopoulos2011inducible} that,
the cancer cell populations starting from different initial proportions
can return towards equilibrium proportions over time.
The arise of this \emph{phenotypic equilibrium} has performed as a profound indication
in support of the intrinsic homeostasis in cancer.
To explain this phenomenon, we investigate the dynamics of phenotypic proportions
in this section.

The model in previous section describes the population dynamics of the absolute cell numbers of different phenotypes. However in reality,
relative numbers, \emph{i.e.} the proportions of cell phenotypes are usually measured by fluorescence-activated cell sorting (FACS)
experiments. Thus one often converts the population model to a frequency model.

Let $N_t$ be the total number of the population, $N_t=X^0_t+X^1_t+...+X^m_t$, the proportion of $X^i_t$ is defined as
$x^i_t=X^i_t/N_t.$
Then $\overrightarrow{x_t}=(x^0_t, x^1_t,..., x^m_t)^T$ is the vector describing the phenotypic proportions.
Let $\langle \overrightarrow{x_t} \rangle$ be the expectation of $\overrightarrow{x_t}$, from Eq. (\ref{ODE}) we can obtain
the nonlinear (second order) ODEs governing the dynamics of $\langle \overrightarrow{x_t} \rangle$ as follows
(see mathematical details in \ref{appendixA}):
\begin{eqnarray}
\frac{d \langle\overrightarrow{x_t}\rangle}{d t}=A\langle\overrightarrow{x_t}\rangle-\langle\overrightarrow{x_t}\rangle e^TA\langle\overrightarrow{x_t}\rangle,
\label{nonlinearODE}
\end{eqnarray}
where $e=(1,1,...,1)^T.$ Note that $x^0_t+x^1_t+...+x^m_t=1$, if replacing $x^m_t$ by $1-(x^0_t+x^1_t...+x^{m-1}_t)$,
the dimension of Eq. (\ref{nonlinearODE}) should reduce from $m+1$ to $m$ (see \ref{appendixA}).

To show the stability of the frequency model Eq. (\ref{nonlinearODE}), we have Theorem \ref{Thm1}:

\newtheorem{theorem}{Theorem}
\begin{theorem}
\textbf{If the Perron-Frobenius eigenvalue $\lambda_1$ of the matrix $A$ in Eq. (\ref{matrix}) is simple (i.e.
the algebraic multiplicity of $\lambda_1$ is equal to one), then Eq. (\ref{nonlinearODE})
has one and only one stable fixed point and no stable limit cycle. \\}
(Technically, in very few cases it is necessary to add a small perturbation to the initial state for
completing the final proof, see proof in \ref{appendixB})
\label{Thm1}
\end{theorem}
The Perron-Frobenius eigenvalue $\lambda_1$ is the largest real eigenvalue
satisfying that Re$\lambda<\lambda_1$ for every other $\lambda$ (\ref{appendixB}).
On account of the inevitability of perturbations in real world,
it can be proved that almost surely the algebraic multiplicity of $\lambda_1$ is equal to one (\ref{appendixC}).
Especially when validating the model with biological experiments, the condition of Theorem \ref{Thm1}
can easily be satisfied.
Theorem \ref{Thm1} thus theoretically explains the phenotypic equilibrium phenomena observed in cancer cell lines,
that is, starting from different initial proportions, the cancer cell population
can return towards stable equilibrium proportions as time passes.
Furthermore, note that Theorem \ref{Thm1} holds for the general cases with any finite phenotypes,
it can be predicted that the phenotypic equilibrium should be universal in multi-phenotypic
population of cancer cells.

In next section we will turn our attention from long-term asymptotic behavior to transient dynamics of the model.

\section{Transient analysis and cell plasticity}

In this section, we investigate how the cell plasticity influences the transient proportions of cell phenotypes and
maintains the heterogeneity of cancer.

To explore the proportion of any phenotype $i$ ($0\leq i \leq m$),
from Eq. (\ref{nonlinearODE}) we have
\begin{align}
\frac{d\langle x^i_t\rangle}{dt}
=\sum_{j=0}^{m}(a_{ij}\langle x^j_t\rangle)-\langle x^i_t\rangle\sum_{n=0}^{m}\sum_{j=0}^{m}(a_{nj}\langle x^j_t\rangle).
\label{transient}
\end{align}
We are interested in how $\langle x^i_t\rangle$ changes shortly after the drastic reduction of it.
This problem is of particular interest to cancer biologists because of two major lines: One is that in cell culture experiments,
they care about the transient changes of cancer cells shortly after the cell sorting. The other is that they
are concerned about how the residual cancer cells relapse after the treatments.

Assume that the proportion of phenotype $i$ at time $t_0$ becomes very small,  i.e. $\langle x^i_{t_0}\rangle\approx 0$.
Then Eq. (\ref{transient}) can be approximately simplified as follows
\begin{equation}
\left.\frac{d\langle x^i_t\rangle}{dt}\right|_{t=t_0}\approx \sum_{j\neq i}^{m}(a_{ij}\langle x^j_{t_{0}}\rangle).
\label{initialrate}
\end{equation}
Eq. (\ref{initialrate})
indicates that the transient growth of $\langle x^i_{t}\rangle$ at time $t_0$
is almost contributed by the cell-state conversions
from other phenotypes provided that current proportion of phenotype $i$ is very small.
That is, as soon as the cancer cells of one phenotype dramatically decreases, cancer cell plasticity
helps increase their phenotypic proportion, which serves as a mutual-helping mechanism in transient dynamics.
In particular, suppose the great majority of CSCs have already been eliminated (\emph{e.g.} by CSCs-targeted
drugs or cell sorting), \emph{i.e.} $\langle x^0_{t_0}\rangle\approx 0$. It is easy to distinguish
the conventional CSC model from the one with cell plasticity by comparing their
transient dynamics:
In conventional CSC model without cancer cell plasticity,
the growth of CSCs relies only on the self-renewal of CSCs themselves, the initial growth
of CSCs proportion should be very limited (constrained by the rate limitation of cell division cycle);
In contrast, the de-differentiations from other NSCCs can effectively
speed up the transient grow rate of CSCs proportion. In this way, there should be a disparity of
the transient increase between these two models shortly after the initial time $t_0$.
To further illustrate the impact of cell plasticity on transient change of CSCs proportion,
we will discuss two concrete examples in next two subsections: CSC-NSCC model and S-B-L model.

\subsection{CSC-NSCC model}
\label{}

CSC-NSCC model is the most widely studied cell plasticity model. In this model,
besides CSCs, all the other non-CSCs are grouped into one whole phenotype.
We now treat the CSC-NSCC model as a two-phenotypic example of the multi-phenotypic
framework, which contains six cellular processes as follows:\\
1) CSC $\overset{\alpha_{00}}{\longrightarrow}$ CSC+CSC;\\
2) CSC $\overset{\alpha_{01}}{\longrightarrow}$ CSC+NSCC;\\
3) CSC $\overset{\alpha_{0}}{\longrightarrow}$ $\emptyset$.\\
4) NSCC $\overset{\alpha_{10}}{\longrightarrow}$ CSC.\\
5) NSCC $\overset{\alpha_{11}}{\longrightarrow}$ NSCC+NSCC;\\
6) NSCC $\overset{\alpha_{1}}{\longrightarrow}$ $\emptyset$;\\

Let $s_t$ be CSCs proportion, and $n_t=1-s_t$ be NSCCs proportion, the equation governing the change of CSCs proportion is then
given by
\begin{equation}
\frac{d\langle s_t \rangle}{dt}=-C\langle s_t \rangle^2+D\langle s_t \rangle+\alpha_{10},
\label{CSC-NSCC}
\end{equation}
where $C=(\alpha_{00}+\alpha_{01}-\alpha_{0})-(\alpha_{11}-\alpha_{1})$ and $D=(\alpha_{00}-\alpha_0)-(\alpha_{11}+\alpha_{10}-\alpha_1)$. Note that
$\alpha_{10}$ is the transition rate from NSCCs to CSCs, the model will reduce to
the conventional CSC model without cell plasticity by letting $\alpha_{10}=0$.
Fig. 2 shows the comparison of the models with and without cell plasticity
by validating them to the published data on SW620 colon cancer cell line \cite{yang2012dynamic}. The method for parameter fitting
we performed is presented in \ref{appendixD1}. It is found that
the major difference between the two models lies in purified NSCCs case (Fig. 2A): Starting from very small CSCs proportion ($0.6\%$ purity),
the initial growth of CSCs proportion predicted by the model without cell plasticity is very slow, the growth rate will gradually
increase until CSCs proportion tends to its equilibrium level, which shows a typical sigmoidal growth pattern \cite{rodriguez2013tumor}.
However, the model with cell plasticity predicted a \emph{transient increase} of CSCs proportion shortly after the initiation,
which is in line with the experimental data. Even though both models predict the final phenotypic equilibrium, only can the model with
cell plasticity fit the transient change on the data. This indicates that cancer cell plasticity could play more essential role
in transient dynamics than in long-term behavior of CSCs proportion.
A similar result is obtained if the stochastic version of the model is simulated using Gillespie algorithm (see \ref{appendixD}).

\subsection{S-B-L model}
\label{S-B-L}

We now consider a three-phenotypic example of our model.
Enlightened by Gupta \emph{et al.}'s work \cite{gupta2011stochastic},
the three-phenotypic model comprises of stem-like (S), basal (B) and luminal
cells (L). Based on our framework, S-B-L model contains twelve reactions as follows:\\
1) S $\overset{\alpha_{00}}{\longrightarrow}$ S+S;\\
2) S $\overset{\alpha_{01}}{\longrightarrow}$ S+B;\\
3) S $\overset{\alpha_{02}}{\longrightarrow}$ S+L;\\
4) S $\overset{\alpha_{0}}{\longrightarrow}$ $\emptyset$.\\
5) B $\overset{\alpha_{11}}{\longrightarrow}$ B+B;\\
6) B $\overset{\alpha_{10}}{\longrightarrow}$ S;\\
7) B $\overset{\alpha_{12}}{\longrightarrow}$ L;\\
8) B $\overset{\alpha_{1}}{\longrightarrow}$ $\emptyset$;\\
9) L $\overset{\alpha_{22}}{\longrightarrow}$ L+L;\\
10) L $\overset{\alpha_{20}}{\longrightarrow}$ S;\\
11) L $\overset{\alpha_{21}}{\longrightarrow}$ B.\\
12) L $\overset{\alpha_{2}}{\longrightarrow}$ $\emptyset$;\\

Let $(s_t, b_t, l_t)^T$ be the vector describing the proportions of (S, B, L), then we have
  \begin{equation}
  \left\{
   \begin{aligned}
   \frac{d\langle s_t \rangle}{dt}&=(\alpha_{00}-\alpha_{0})\langle s_t \rangle+\alpha_{10} \langle b_t \rangle+\alpha_{20}\langle l_t \rangle-s_t[A_1\langle s_t \rangle+A_2\langle b_t \rangle+A_3\langle l_t \rangle]\\
   \frac{d\langle b_t \rangle}{dt}&=\alpha_{01}\langle s_t \rangle+(\alpha_{11}-\alpha_{10}-\alpha_{12}-\alpha_{1})\langle b_t \rangle+\alpha_{21}\langle l_t \rangle-\langle b_t \rangle[A_1\langle s_t \rangle+A_2\langle b_t \rangle+A_3\langle l_t \rangle]\\
   \frac{d\langle l_t \rangle}{dt}&=\alpha_{02}\langle s_t \rangle+\alpha_{12}\langle b_t \rangle+(\alpha_{22}-\alpha_{20}-\alpha_{21}-\alpha_{2})\langle l_t \rangle-\langle l_t \rangle[A_1\langle s_t \rangle+A_2\langle b_t \rangle+A_3\langle l_t \rangle]\\
   \end{aligned}
   \right.
   \label{S-B-L}
  \end{equation}
where $A_1=\alpha_{00}+\alpha_{01}+\alpha_{02}-\alpha_0$, $A_2=\alpha_{11}-\alpha_1$, $A_3=\alpha_{22}-\alpha_2$.
Fig. 3 shows the predictions of S-B-L model by fitting to cell-state dynamics on data of SUM159 breast cancer cell line
\cite{gupta2011stochastic}(see methods in \ref{appendixD1}). Starting from very small $\langle s_0 \rangle$, in contrast to the gradual increase of $\langle s_t \rangle$
predicted by the model with $\alpha_{20}=0$, there is an \emph{overshooting} of $\langle s_t \rangle$ predicted by the model with $\alpha_{20}>0$,
which is in line with the cell-state dynamics observed in experiment.
That is, the proportion of stem-like cells is rapidly elevated by de-differentiation to the value above the equilibrium level, and then returns towards the final equilibrium as time passes. It has been reported that CSC-NSCC model can never perform overshooting \cite{zhoupopulation, jia2013overshoot}, which implies that the overshooting phenomenon are rooted in the diversity of phenotype in the multi-phenotypic model. Moreover,
note that overshooting behavior is a result of biological response to environmental stimulus through evolution \cite{ma2009defining}, cancer cell plasticity may have meaningful implications for the evolution of cancer.
One possible explanation is that the reversible phenotype switching between cells provides an advantageous cooperative strategy for cancer cell populations during their competitions with various anticancer mechanisms \cite{axelrod2006evolution}, thus increases their collective fitness. This idea is in line with the literature in ecology and population genetics \cite{kussell2005phenotypic,acar2008stochastic} regarding phenotypic variability as a surviving strategy for biological populations in fluctuating environments.

\section{Conclusions}
\label{}

Instead of the hierarchical structure of cancer cells proposed by cancer stem cell theory, the updated CSC paradigm with cell plasticity
suggests a reversible nature between different cancer cell phenotypes. In this study we proposed a generalized multi-phenotypic cancer model
for providing a mathematical framework to investigate the interplay between cell plasticity and diversity of phenotypes in cancer.
Based on our model, some interesting and insightful results have been achieved.
On one hand, it was shown (in Theorem \ref{Thm1}) that the phenotypic equilibrium phenomena can be predicted
by the stability of our frequency model.
On the other hand, by comparing the models with and without cell plasticity, it was found that
cell plasticity may play more essential role in transient dynamics. It has been shown that
only can the models with cell plasticity predict the transient increase and overshooting phenomena in CSCs proportion.
In particular, the overshooting phenomenon performed by the S-B-L model was shown to be one of the salient features
in the multi-phenotypic models with cell plasticity.
With the further study of the multi-phenotypic models with cell plasticity,
more interesting and insightful phenomena would be found in future.

Cancer cell plasticity has enriched the theory of cancer stem cell and comes along with new fundamental problems
in cancer biology: Firstly, its molecular mechanism is poorly understood. It was reported that TGF-$\beta$ might have important roles
in the process of cell-state conversions through activating epithelial mesenchymal transition (EMT) \cite{yang2012dynamic}.
Further studies on related molecular models should be important tasks. Accordingly,
besides the population-level models studied in this work, molecular-level models should also be accounted for in future.
Secondly, even though
cell plasticity is found to be intimately related to the heterogeneity of cancer,
cancer complexity is a result of multifactorial mechanisms (\emph{e.g.} clonal evolution, tumor microenvironment, reversible cell plasticity and etc \cite{meacham2013tumour}),
and it is not clear in which cancer and to what extent the disease progression arises from the cell plasticity.
That is, how to distinguish cell plasticity from other sources of cancer heterogeneity should be a central problem in future, for both theoretical and experimental researchers.

\section*{Acknowledgements}
We thank the referees for constructive comments.
We also thank Prof. Minping Qian, Drs. Hao Ge, Chen Jia,  Linyuan Liu, Hong Qian,
Xin Tong, Benjamin Werner, Zhen Xie, Gen Yang, Michael Q Zhang, and Ruixiang Zhang,
for helpful discussions.
B. W. greatly acknowledges the generous sponsorship from Max-Planck Society.
This work is supported by the Fundamental Research Funds for the Central Universities in China.

\section{Appendix}

\subsection{Derivation of Eq. (\ref{nonlinearODE})}
\label{appendixA}

For Eq. (\ref{ODE})
$$\frac{d\langle X^i_t\rangle}{dt}=a_{i0}\langle X^0_t\rangle+a_{i1}\langle X^1_t\rangle+...+a_{im}\langle X^{m}_t\rangle.$$
Note that
$$x^i_t=\frac{X^i_t}{X^0_t+X^1_t+...+X^m_t}=\frac{X^i_t}{N_t},$$
$$\frac{d\langle X^i_t\rangle}{dt}=\frac{d(\langle x^i_t\rangle\langle N_t\rangle)}{dt}=\langle x^i_t\rangle\frac{d}{dt}\langle N_t\rangle
+\langle N_t\rangle\frac{d\langle x^i_t\rangle}{dt},$$
then
\begin{align*}
\frac{d\langle x^i_t\rangle}{dt}
&=\frac{1}{\langle N_t\rangle}\frac{d\langle X^i_t\rangle}{dt}-\frac{\langle x^i_t\rangle}{\langle N_t\rangle}\frac{d\langle N_t\rangle}{dt}\\
&=\sum_{j=0}^{m}(a_{ij}\langle x^j_t\rangle)-\langle x^i_t\rangle\sum_{n=0}^{m}\sum_{j=0}^{m}(a_{nj}\langle x^j_t\rangle).
\end{align*}
Then we have Eq. (\ref{nonlinearODE}).

By letting $x^m_t=1-(x^0_t+x^1_t...+x^{m-1}_t)$, we have (for $0\leq i \leq m-1$)
\begin{align*}
\frac{d\langle x^i_t\rangle}{dt}
&=\sum_{j=0}^{m}(a_{ij}\langle x^j_t\rangle)-\langle x^i_t\rangle\sum_{n=0}^{m}\sum_{j=0}^{m}(a_{nj}\langle x^j_t\rangle)\\
&=\sum_{j=0}^{m-1}(a_{ij}\langle x^j_t\rangle)+a_{im}\langle x^m_t\rangle-\langle x^i_t\rangle\sum_{n=0}^{m}\left(\sum_{j=0}^{m-1}(a_{nj}\langle x^j_t\rangle)+a_{nm}\langle x^m_t\rangle\right)\\
&=\sum_{j=0}^{m-1}(a_{ij}-a_{im})\langle x^j_t\rangle-\langle x^i_t\rangle\sum_{n=0}^{m}\left(\sum_{j=0}^{m-1}(a_{nj}-a_{nm})\langle x^j_t\rangle+a_{nm}\right)+a_{im}\\
&=\underbrace{-\langle x^i_t\rangle\sum_{n=0}^{m}\sum_{j=0}^{m-1}(a_{nj}-a_{nm})\langle x^j_t\rangle}_{\textrm{nonlinear term}}
\underbrace{-\langle x^i_t\rangle\sum_{n=0}^{m}a_{nm}+\sum_{j=0}^{m-1}(a_{ij}-a_{im})\langle x^j_t\rangle}_{\textrm{linear term}}
\underbrace{+a_{im}.}_{\textrm{constant term}}\\
\end{align*}

\subsection{Proof of Theorem \ref{Thm1}}
\label{appendixB}

Let us first consider the properties of Eq. (\ref{ODE}).
We will show that there exists a largest real eigenvalue $\lambda_1$ of $A$
satisfying Re$\lambda<\lambda_1$ for every other $\lambda$.
Note that the off-diagonal elements of the matrix $A$ are non-negative,
$A+\kappa I$ becomes a non-negative matrix ($I$ is the identity matrix) when $\kappa$ is large enough.
By Perron-Frobenius theory (see chapter 1 in \cite{seneta1981non}), $A+\kappa I$ has an real eigenvalue $\lambda'_1$
satisfying $Re\lambda'<\lambda'_1$ for any other eigenvalue $\lambda'$ of $A+\kappa I$.
It is easy to show that $\lambda'_1-\kappa$ is just the largest real eigenvalue (called \emph{Perron-Frobenius eigenvalue} $\lambda_1$)
satisfying that Re$\lambda<\lambda_1$ for every other $\lambda$ of $A$
\footnote{If $\lambda'$ is an eigenvalue of $A+\kappa I$, then $det|A+\kappa I-\lambda' I|=0$. For matrix $A$, we have $det|A-(\lambda'-\kappa)I|=0$, namely $\lambda' -\kappa$ is an eigenvalue of $A$. Then $A$ has an real eigenvalue $\lambda'_1-\kappa$ satisfying $Re(\lambda'-\kappa)<\lambda'_1-\kappa$ for any other eigenvalue $\lambda'-\kappa$ of $A$.}
.

Suppose
$\lambda_1$ is simple, the solution of Eq. (\ref{ODE}) can be expressed as
\begin{equation*}
\begin{split}
&\langle\overrightarrow{X_t}\rangle=c_{1,1}\vec{u}e^{\lambda_{1}t}+\sum_{j=2}^{n}\sum_{l=1}^{n_j}c_{j,l}\sum_{i=1}^{n_j}\vec{r^{j}_{l,i}}t^{i-1}e^{\lambda_{j}t},
\end{split}
\end{equation*}
where $\lambda_{1},\lambda_{2},\cdots{}\lambda_{n}$ are the different eigenvalues of $A$, $n_j$ is the algebraic multiplicity of $\lambda_{j}$, $\vec{u}$ is the normalized (\emph{i.e.} $u_0+u_2+...+u_m=1$) right eigenvector of $\lambda_1$, $\vec{r^j_{l,i}}$ is the corresponding eigenvector of $\lambda_{j}$, $c_{j,l}$ is a constant determined by initial values. Note that Re$\lambda_i<\lambda_1~(i\neq 1)$, when $c_{1,1}\neq{}0$,
\begin{equation*}
\begin{split}
&\frac{\langle\overrightarrow{X_t}\rangle}{c_{1,1}e^{\lambda_{1}t}}=\vec{u}+\sum_{j=2}^{n}\sum_{l=1}^{n_j}\frac{c_{j,l}}{c_{1,1}}\sum_{i=1}^{n_j}\vec{r^{j}_{l,i}}t^{i-1}e^{(\lambda_{j}-\lambda_1)t}\rightarrow \vec{u}.
\end{split}
\end{equation*}
Thus
$$\langle\overrightarrow{x_t}\rangle=\frac{\langle\overrightarrow{X_t}\rangle}{\langle N_t \rangle}
=\frac{\overrightarrow{X_t}/c_{1,1}e^{\lambda_{1}t}}{\langle N_t \rangle/c_{1,1}e^{\lambda_{1}t}}
\rightarrow \frac{\vec{u}}{u_0+u_2+...+u_m}$$
Since $\vec{u}$ has been normalized, $\langle\overrightarrow{x_t}\rangle$
will tend to $\vec{u}$ when $t\rightarrow{}+\infty{}$.

In the following we will handle the case of $c_{1,1}=0$.
Let $t=0$, we get the linear equations of $c_{j,l}$:
\begin{equation*}
\begin{split}
&c_{1,1}\vec{u}+\sum_{j=2}^{n}\sum_{l=1}^{n_j}c_{j,l}\vec{r^{j}_{l,1}}=\langle\overrightarrow{X_0}\rangle^T
\end{split}
\end{equation*}
Denote the coefficient matrix by $B=[\vec{u}\ \vec{r^{2}_{1,1}}\ \vec{r^{2}_{2,1}}\cdots \vec{r^{n}_{n_n,1}}]$, and let $B^*$ be $B$ with its first column replaced by $\langle\overrightarrow{X_0}\rangle^T$. Notice that columns of $B$ are linear independent. Then from Cramer's rule
\begin{equation*}
\begin{split}
&c_{1,1}=\frac{det|B^*|}{det|B|},
\end{split}
\end{equation*}

When $c_{1,1}=0$, namely $det|B^*|=0$, we can find an $n$-dimensional vector $\vec{v}$ which is linear independent with the last $n-1$ columns of $B$. Actually, with a small perturbation $\varepsilon{}\vec{v}$ to $\langle\overrightarrow{X_0}\rangle^T$, all the columns of $B^*$ are linear independent, and $c_{1,1}\neq{}0$. Note that fluctuations are inevitable in real world, $c_{1,1}\neq0$ always holds in reality.

\subsection{The algebraic multiplicity of Perron-Frobenius eigenvalue}
\label{appendixC}

It was shown above that the Perron-Frobenius eigenvalue $\lambda_1$ being \emph{simple} is essential for proving Theorem \ref{Thm1}.
Here we will show that this condition is easily satisfied in real world. In fact, a more general result can be obtained as follows:

\newtheorem{proposition}{Proposition}
\begin{proposition}
Denote $\mathbb{R}^{n\times n}$ as the space of $n\times n$ real square matrices, equipped with Lebesgue measure.
Then its subset of the matrices with repeated eigenvalues has measure zero.
\label{Thm2}
\end{proposition}

To prove this proposition, we need following lemma first.

\newtheorem{lemma}{Lemma}
\begin{lemma}
 For any non-zero polynomial $f(x_1, x_2, \cdots ,x_m)$ $(m\geq 1)$, its zero set \footnote{ The zero set of $f(x_1, x_2, \cdots ,x_m)$
 is defined as $\{(x_1, x_2, \cdots ,x_m)\in \mathbb{R}^m | f(x_1, x_2, \cdots ,x_m)=0\}$}
 has measure $0$ (w. r. t. Lebesgue measure on $\mathbb{R}^m $).
\label{lemma}
\end{lemma}

\textbf{Proof of Lemma \ref{lemma}}:  We use inductive method. For $m=1$, the polynomial has only finite roots, whose zero set has measure $0$. Assume we have proved the lemma for $m=k-1$. For $m=k$, let us take $f(x_1, x_2, \cdots ,x_k)$ as the polynomial of $(x_2, \cdots ,x_k)$, then the coefficients can be regarded as polynomials of $x_1$. It is easy to see that at most finite $x_1\in \mathbb{R}$ can make all coefficients equal to $0$. Denote such set of $x_1$ by $S$, then $\{(x_1, x_2, \cdots ,x_k|\ x_1\in S\}$ has measure $0$ in $\mathbb{R}^k$. For any fixed $x_1\notin S$, $f(x_1, x_2, \cdots ,x_k)$ is a non-zero polynomial of $(x_2, \cdots ,x_k)$. From inductive assumption we know its zero set has measure $0$ in $\mathbb{R}^{k-1}$. By Fubini-Tonelli theorem, we know that $\{(x_1, x_2, \cdots ,x_k)|\ x_1\notin S,\ f(x_1, x_2, \cdots ,x_k)=0\}$ has measure $0$ in $\mathbb{R}^k$. Thus $\{(x_1, x_2, \cdots ,x_k)|\ f(x_1, x_2, \cdots ,x_k)=0\}$ has measure $0$ in $\mathbb{R}^k$.

\textbf{Proof of Proposition \ref{Thm2}}: It is easy to show that for any square matrix in $\mathbb{R}^{n\times n}$, the coefficients of its characteristic polynomial are polynomials of matrix entries. Thus the discriminant of the characteristic polynomial
\footnote{The discriminant of $m$-order polynomial $a_m x^m+a_{m-1}x^{m-1}\cdots+a_1 x+a_0$ is a polynomial function (see pp. 16-19 in \cite{dickenstein2005solving}) for concrete form) of its coefficients $a_m, a_{m-1}, \cdots a_1, a_0$. A polynomial has repeated root if and only if its discriminant is $0$ \cite{dickenstein2005solving}. For example, the discriminant of quadratic polynomial $ax^2+bx+c$ is $b^2-4ac$. $ax^2+bx+c$ has repeated root if and only if $b^2-4ac=0$.}
is again a polynomial of matrix entries. Applying Lemma \ref{lemma} to the discriminant of characteristic polynomial, we have
that the zero set of the discriminant has measure zero. And then according to the fact that a polynomial has repeated root if and only if its discriminant is $0$ \cite{dickenstein2005solving}, the set of the matrices with repeated eigenvalues has measure zero. The proof is completed.

Proposition \ref{Thm2} implies that almost surely the Perron-Frobenius eigenvalue $\lambda_1$ should be simple.
In particular, perturbations are inevitable in the real work, so $\lambda_1$ can easily be simple.
To further illustrate this result, we consider a $3\times 3$ example:
\begin{equation}
A=\left(\begin{array}{ccc}
3 & 2 & 3 \\
0 & 3 & 5 \\
0 & 0 & 2 \\
\end{array}\right),
\label{matrix3}
\end{equation}
its eigenvalues are $\lambda_1=\lambda_2=3$, $\lambda_3=2$. In this case the Perron-Frobenius eigenvalue $\lambda_1$
is repeated. However, if adding perturbations to $A$, \emph{e.g.}
\begin{equation}
A=\left(\begin{array}{ccc}
3.23 & 2.19 & 3.22 \\
0 & 3.78 & 5.11 \\
0 & 0 & 2.31 \\
\end{array}\right),
\label{matrix4}
\end{equation}
its eigenvalues become $\lambda_1=3.78$, $\lambda_2=3.23$, $\lambda_3=2.31$. In this way the Perron-Frobenius eigenvalue $\lambda_1$ become simple.

Since the perturbations from experimental measurements and natural fluctuations are inevitable in reality, the Perron-Frobenius eigenvalue $\lambda_1$ in $A$ is seldom to be repeated. For example, consider the S-B-L model in Sec 4.2, let
$\alpha_{00}=0.34$, $\alpha_{01}=0.29$, $\alpha_{02}=0.34$, $\alpha_{0}=0.11$,
$\alpha_{11}=0.43$, $\alpha_{10}=0.07$, $\alpha_{12}=0.13$, $\alpha_{1}=0.09$,
$\alpha_{22}=0.36$, $\alpha_{20}=0.06$, $\alpha_{21}=0.17$, $\alpha_{2}=0.10$,
then the matrix $A$ turns to
\begin{equation}
A=\left(\begin{array}{ccc}
0.23 & 0.09 & 0.10 \\
0.29 & 0.14 & 0.06 \\
0.34 & 0.07 & 0.03 \\
\end{array}\right).
\label{matrix4}
\end{equation}
The eigenvalues of $A$ are $\lambda_1=0.4391$, $\lambda_2=-0.0844$, $\lambda_3=0.0453$. The Perron-Frobenius eigenvalue $\lambda_1$ is simple.

\subsection{Methods for parameter fitting}
\label{appendixD1}

We used least squares method to estimate the parameters. Let $\vec{y}=(y_0,...,y_n)^T$ be the data on $n$ time points,
$\vec{x}(\vec{\theta})=(x_0,...,x_n)^T$ be the discrete trajectory of the model with parameter vector $\vec{\theta}$.
The best-fitting parameters are obtained
by solving the following least squares problem
$$\min_{\vec{\theta}}\sum_{i=0}^{n}(x_i-y_i)^2.$$
Specifically, we first discuss the parameter fitting of the CSC-NSCC model to the data on SW620 cell line \cite{yang2012dynamic}.
There are four groups of data in their experiment, each group contains 13 time points (from day 0 to day 24, the data points were recorded every two days).
That is, there are a total of $13\times4=52$ time points.
We estimated the parameters in Eq. (\ref{CSC-NSCC}) by fitting to these 52 data points of four groups all together.
It should be noted that,
based on population-level data measured by FACS experiment, only can we estimate the values of
$C$, $D$ and $\alpha_{10}$ in Eq. (\ref{CSC-NSCC}), we cannot estimate all the $\alpha_{ij}$ or ${\alpha_i}$
individually. However, it is interesting that $\alpha_{10}$ is presented
as an independent coefficient, so we can still compare the models with and without cell plasticity
based on the population-level analysis. We list the main points for parameter fitting as follows:
\begin{enumerate}
  \item \textbf{Fit four groups data together:} Let $y^{j}_{i}$ be the $i^{th}$ time point in $j^{th}$ groups, then we need to solve the following least squares problem
  $$\min_{(C,D,\alpha_{10})}\sum_{j=1}^4\sum_{i=0}^{12}(x^j_i-y^j_i)^2.$$
  \item \textbf{Fixed initial states:} We set the initial states of Eq. (\ref{CSC-NSCC}) are the same as those on data, that is, let $x^j_0=y^j_0$ $(1\leq j\leq 4)$.
  \item \textbf{Parameter range:} Based on the rate limitation of human cell proliferation cycle \cite{cowan2004derivation}, we assume that
  each cell requires at least one day to finish a cell cycle. To rescale the discrete time point to continuous time and note that
  the data was recorded every two days, we have $0\leq \alpha_{ij}\leq 2\log2$. Meanwhile, for the non-extinction of the cell population, we
  assume that cell death rate is smaller than symmetric division rate, i.e. $0\leq \alpha_{i} < \alpha_{ii}\leq 2\log2$.
\end{enumerate}
We performed \textsf{fmincon} algorithm in \emph{Matlab} to calculate optimal values of $(C^*,D^*,\alpha^*_{10})$.
To test the uniqueness of the solutions,
100 different starting values of $(C^0,D^0,\alpha^0_{10})$ were uniformly selected on the basis of the above parameter range.
It was shown that the values of $(C,D,\alpha_{10})$ converge to $C^*\sim(10^{-6}, 10^{-5})$, $D^*=-0.4002\pm 0.0001$, and $\alpha^*_{10}=0.2630\pm0.0002$.
For the goodness of fit, the coefficient of determination is used to
measure how well the observations on data are predicted by the models, that is,
$$R^2=1-\frac{SS_{res}}{SS_{tot}}=1-\frac{0.0588}{0.9921}=0.9407,$$
where $SS_{res}$ is the sum of squared residuals and $SS_{tot}$ is the total sum of squares (proportional to the sample variance) of data.
By letting $\alpha_{10}=0$, with the same numerical scheme we have $C^*=1.3865\pm0.0002$, $D^*=0.9550\pm 0.0001$, and
the coefficient of determination is
$$R^2=1-\frac{SS_{res}}{SS_{tot}}=1-\frac{0.5932}{0.9921}=0.4021.$$
Fig 2A shows that the disparity between the predictions by the two models mainly lies in the transient dynamics in purified NSCCs case,
where only can the model with $\alpha_{10}>0$ predict the transient increase of CSCs proportion.

Now we discuss the parameter fitting of the S-B-L model. The time-series data we used is the cell-state dynamics on SUM159 breast cancer cell line (Figure 3 in \cite{gupta2011stochastic}
\footnote{It should be noted that, this data was produced by discrete-time Markov chain with transition matrix
\begin{equation}
P=\left(\begin{array}{ccc}
0.58 & 0.35 & 0.07 \\
0.01 & 0.99 & 0 \\
0.04 & 0.49 & 0.47 \\
\end{array}\right).
\label{MarkovChain}
\end{equation}
According to the methods in \cite{gupta2011stochastic}
this cell-state dynamic data was in good line with experimental results on SUM159 breast cancer cell line.}).
There were three cases in their experiments (purified stem-like cells case, purified basal cells case and purified luminal cells case).
In each case, 12 time points were recorded for each cell phenotype. That is, a total of $3\times 12\times 3=108$ data points were
recorded. Note that $s_t+b_t+l_t=1$, the effective number of the data is actually reduced to 72. By letting $l_t=1-s_t-b_t$, we have
  \begin{equation}
  \left\{
   \begin{aligned}
   \frac{d\langle s_t \rangle}{dt}&=A_4\langle s_t \rangle+\alpha_{10} \langle b_t \rangle+\alpha_{20}\langle l_t \rangle-s_t[A_1\langle s_t \rangle+A_2\langle b_t \rangle+A_3\langle l_t \rangle]\\
   \frac{d\langle b_t \rangle}{dt}&=\alpha_{01}\langle s_t \rangle+A_5\langle b_t \rangle+\alpha_{21}\langle l_t \rangle-\langle b_t \rangle[A_1\langle s_t \rangle+A_2\langle b_t \rangle+A_3\langle l_t \rangle]\\
   \langle l_t \rangle &=1-s_t-b_t\\
   \end{aligned}
   \right.
   \label{S-B-L2}
  \end{equation}
where $A_4=\alpha_{00}-\alpha_{0}$, $A_5=\alpha_{11}-\alpha_{10}-\alpha_{12}-\alpha_{1}$.
Similar to the CSC-NSCC model, on the basis of population level data,
we cannot estimate all the twelve parameters $\alpha_{ij}$ and $\alpha_i$ in S-B-L model individually, since
the freedom of the parameters in Eq. (\ref{S-B-L2}) is only nine. Note that the parameters for de-differentiation ($\alpha_{20}$ and $\alpha_{10}$)
are presented as independent coefficients, it is feasible to investigate how de-differentiation influences the proportion of stem-like cells in transient dynamics.
By applying similar numerical scheme we used in the CSC-NSCC model
\footnote{Since the time point in S-B-L model was recorded every day, the parameter constraints are changed to $0\leq \alpha_{ij}\leq \log2$.},
we fitted Eq. (\ref{S-B-L2}) to the cell-state dynamics on data.
Fig. 4 shows the predictions by the model with $\alpha_{20}=0$, and
Fig. 5 shows the predictions by the model with $\alpha_{20}>0$. Note that the cell-state dynamic data in \cite{gupta2011stochastic}
was produced by the Markov chain, it is not surprising that the coefficients of determination of the two models are both close to 1.
But it is note worthy that, the main difference between
the two cases lies in the purified luminal case, where only can the model with $\alpha_{20}>0$ predict the overshooting of
stem-like cells proportion.

\subsection{Stochastic simulations}
\label{appendixD}

The stochastic simulation of CSC-NSCC model is illustrated in Fig. 6. The simulation scheme we used is based on Monte Carlo simulation
for the CME (see Section 11.4.4 in \cite{beard2008chemical}).

\subsection{A remark for the model}
\label{appendixE}

By adding symmetric differentiation of CSC into our model,
\begin{itemize}
  \item CSC $\overset{\beta_{0j}}{\longrightarrow}$ $\textrm{NSCC}_j$+$\textrm{NSCC}_j$ ($1\leq j\leq m$)
\end{itemize}
the dynamics of $\langle\overrightarrow{X_t}\rangle$ can still be formed by the linear ODEs as follows
\begin{eqnarray}
\frac{d \langle\overrightarrow{X_t}\rangle}{d t}=Q\langle\overrightarrow{X_t}\rangle,
\label{ODE2}
\end{eqnarray}
where
\begin{equation}
Q=\left(\begin{smallmatrix}
\alpha_{00}-\alpha_{0}-\sum_{j=1}^m \beta_{0j} & \alpha_{10} & \alpha_{20} & \cdots & \alpha_{m0}\\
\alpha_{01}+2\beta_{01} & \alpha_{11}-\alpha_{1}-\sum_{j\neq 1}\alpha_{1j} & \alpha_{21} & \cdots & \alpha_{m1} \\
\vdots & \vdots & \vdots & \vdots & \vdots  \\
\alpha_{0(m-1)}+2\beta_{0(m-1)} & \cdots & \cdots &  \alpha_{(m-1)(m-1)}-\alpha_{(m-1)}-\sum_{j\neq (m-1)}\alpha_{(m-1)j} & \alpha_{m(m-1)} \\
\alpha_{0m}+2\beta_{0m} & \cdots & \cdots & \alpha_{(m-1)m} & \alpha_{mm}-\alpha_{m}-\sum_{j\neq m}\alpha_{mj} \\
\end{smallmatrix}\right).
\label{matrixQ}
\end{equation}
It is easy to find that the only difference between $Q$ and $A$ in Eq. (\ref{matrix}) lies in the first column,
but this does not change the non-negativity of the off-diagonal elements. Note that the main ingredient of the proof of Theorem \ref{Thm1}
is the non-negativity of the off-diagonal elements in $A$ (see \ref{appendixB}), Theorem \ref{Thm1} still holds for $Q$.

Moreover, the transient properties discussed in our model are still valid for the model with $Q$ here,
by only replacing $a_{ij}$ in Eq. (\ref{transient})
with $q_{ij}$ here. Let us take the CSC-NSCC model as an example.
The equation governing the change rate of CSCs proportion is given by
\begin{equation}
\frac{d\langle s_t \rangle}{dt}=-C^*\langle s_t \rangle^2+D^*\langle s_t \rangle+\alpha_{10},
\label{CSC-NSCC1}
\end{equation}
where $C^*=(\alpha_{00}+\alpha_{01}+\beta_{01}-\alpha_{0})-(\alpha_{11}-\alpha_{1})$ and $D^*=(\alpha_{00}-\alpha_0)-(\alpha_{11}+\alpha_{10}-\alpha_1)$.
the property of transient increase is still dependent on $\alpha_{10}$, which shares the same nature as Eq. (\ref{CSC-NSCC}).






\section*{Figures}
\newpage

\begin{figure}
\begin{center}
\includegraphics[width=1\textwidth]{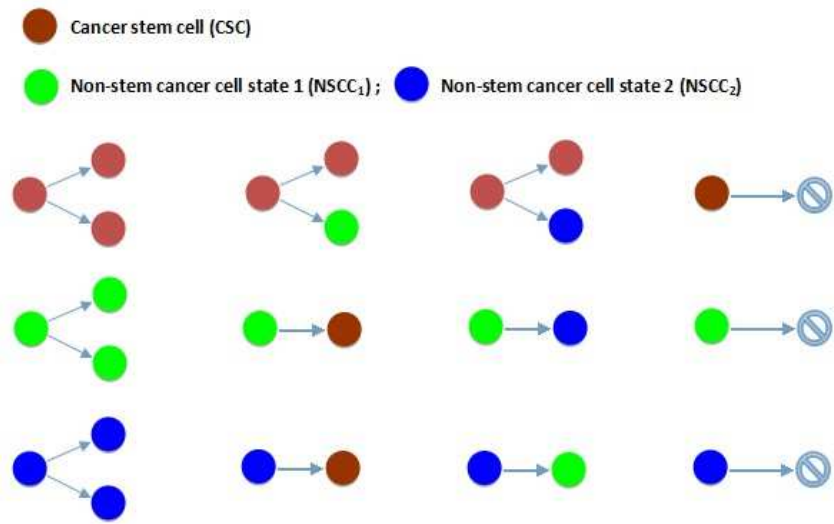}
\caption{The model illustrated by the three-phenotype example (CSC, $\textrm{NSCC}_1$, $\textrm{NSCC}_2$).
For CSC (first row, left to right): symmetric division, asymmetric division (differentiation to $\textrm{NSCC}_1$),
asymmetric division (differentiation to $\textrm{NSCC}_2$), and cell death;
For $\textrm{NSCC}_1$ (second row, left to right): symmetric division, phenotypic conversion to CSC,
phenotypic conversion to $\textrm{NSCC}_2$, and cell death;
For $\textrm{NSCC}_2$ (third row, left to right): symmetric division, phenotypic conversion to CSC,
phenotypic conversion to $\textrm{NSCC}_1$, and cell death.}
\end{center}
\end{figure}

\begin{figure}
\begin{center}
\includegraphics[width=1\textwidth]{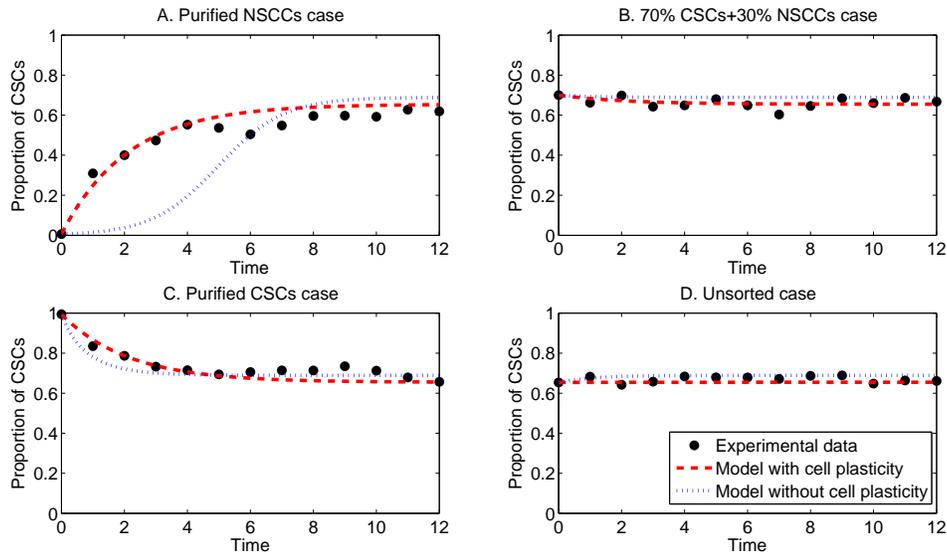}
\caption{The comparison of CSC-NSCC models (\ref{CSC-NSCC})
with and without cell plasticity by fitting to experimental data on
SW620 colon cancer cell line: The black dots are the experimental data in \cite{yang2012dynamic},
the blue dashed lines are the predictions of the model without cell plasticity ($\alpha_{10}=0$),
and the red dashed lines are the predictions of the model with cell plasticity ($\alpha_{10}>0$).
There are four groups of data in the experiment:
A) Purified NSCCs case ($0.6\%$ CSCs+$99.4\%$ NCSSs);
B) $70\%$ CSCs+$30\%$ NCSSs;
C) Unsorted case ($65.4\%$ CSCs+$34.6\%$ NCSSs);
D) Purified CSCs case ($99.4\%$ CSCs+$0.6\%$ NCSSs).
By least squares method, we predicted the values of four
different cases together with the best fitting parameters
(see methods in \ref{appendixD1}).
It is shown that in (B), (C) and (D), the predicted trajectories by both models
are in good accord with the data. In (A), however,
compared to the sigmoidal growth pattern predicted by the model without cell plasticity,
the prediction by the model with cell plasticity shows a transient increase in the first
few days, which is in line with the experimental data.}
\end{center}
\end{figure}

\begin{figure}
\begin{center}
\includegraphics[width=1\textwidth]{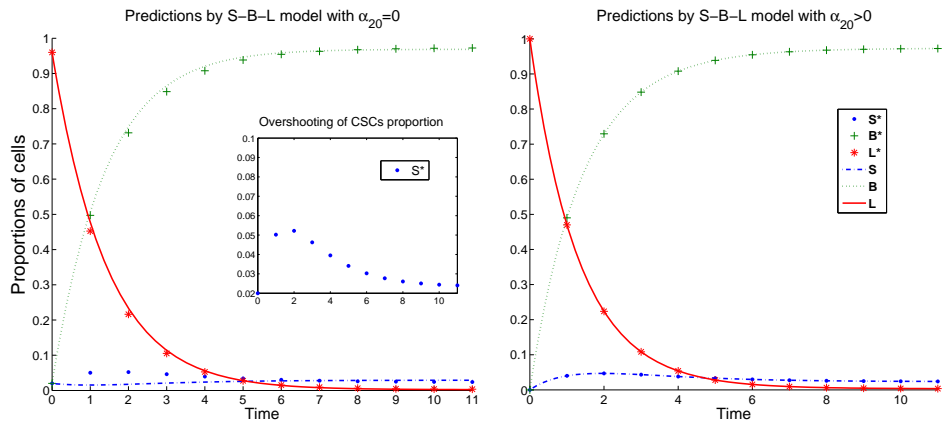}
\caption{Predictions of S-B-L model: We fitted Eq. (\ref{S-B-L})
to the cell-state dynamics on the data of SUM159 breast cancer cell line
(Figure 3 in \cite{gupta2011stochastic}).
$S$, $B$ and $L$ are
the predictions by the model, while $S^*$, $B^*$ and $L^*$
are the cell-state dynamics on data.
There are three groups of data: A) Purified stem-like cells case;
B) Purified basal cells case; C) Purified luminal cells case.
We predicted the time points of three different
cases together with the best fitting parameters.
This figure shows the predictions in purified luminal case,
see \ref{appendixD1} for the other two cases.
The left panel is the predictions by the model without phenotypic conversion from
luminal cells to stem-like cells ($\alpha_{20}=0$), while the right panel is the predictions
by the model with phenotypic conversion from
luminal cells to stem-like cells ($\alpha_{20}>0$). It is shown that only can the model
with cell plasticity predict the overshooting of the proportion of stem-like cells.
The small window in left panel specifically focuses on the overshooting phenomenon of stem-like cells.}
\end{center}
\end{figure}

\begin{figure}
\begin{center}
\includegraphics[width=1\textwidth]{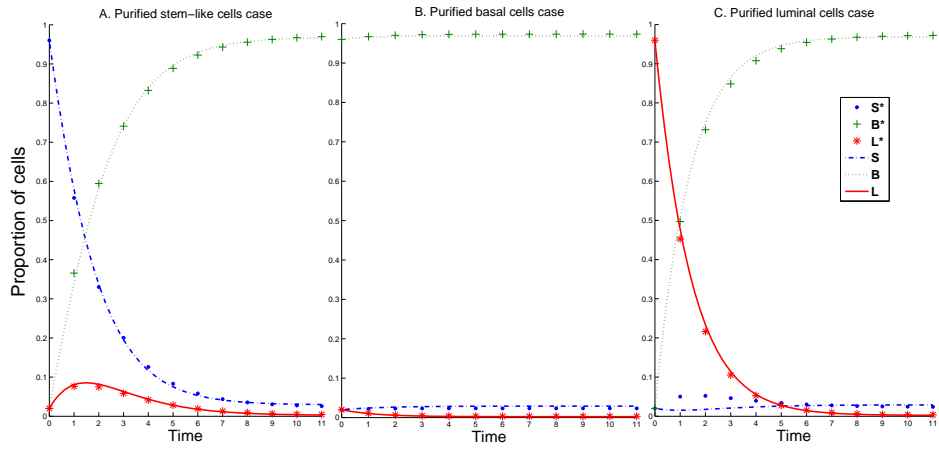}
\caption{Fit Eq. (\ref{S-B-L2}) to cell-state dynamics in \cite{gupta2011stochastic} by letting $\alpha_{20}=0$. The best fitting parameters are $A_4=0.032\pm 0.002$, $\alpha_{10}=0.016\pm 0.003$, $\alpha_{20}=0$, $A_1=0.578\pm 0.002$, $A_2=0.581\pm 0.002$, $A_3=0.550\pm 0.001$, $\alpha_{01}=0.416\pm 0.001$, $A_5=0.567\pm 0.002$ and $\alpha_{21}=0.693\pm 0.001$.
The sum of squared residuals $SS_{res}$ are about 0.06 and the coefficient of determination $R^2=0.9851$. The unfitting part lies in the transient dynamics of the purified luminal cells case.
In particular, the overshooting of the stem-like cells proportion cannot be predicted by the model.}
\end{center}
\end{figure}

\begin{figure}
\begin{center}
\includegraphics[width=1\textwidth]{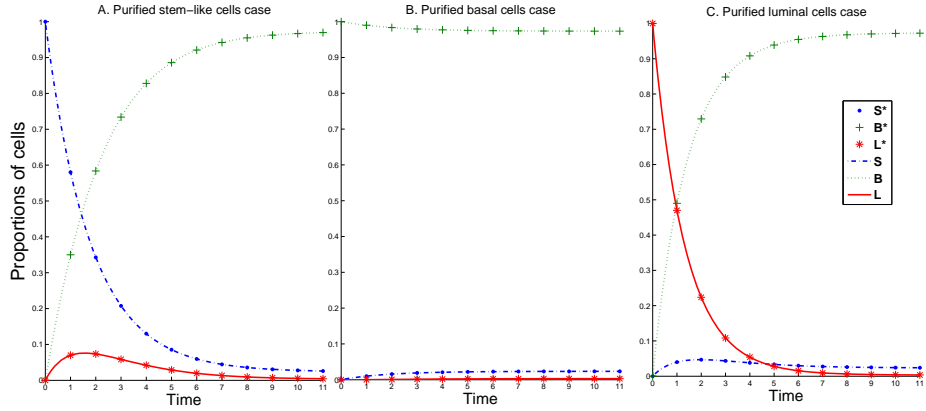}
\caption{Fit Eq. (\ref{S-B-L2}) to cell-state dynamics in \cite{gupta2011stochastic} by letting $\alpha_{20}>0$. The best fitting parameters are $A_4=0.027\pm 0.002$, $\alpha_{10}=0.013\pm 0.001$, $\alpha_{20}=0.071\pm 0.001$, $A_1=0.580\pm 0.002$, $A_2=0.578\pm 0.002$, $A_3=0.579\pm 0.001$, $\alpha_{01}=0.418\pm 0.001$, $A_5=0.567\pm 0.003$, and $\alpha_{21}=0.688\pm 0.001$.
The sum of squared residuals is about $\sim(10^{-7}, 10^{-8})$, so the coefficient of determination $R^2$ is very close to 1.}
\end{center}
\end{figure}

\begin{figure}
\begin{center}
\includegraphics[width=1\textwidth]{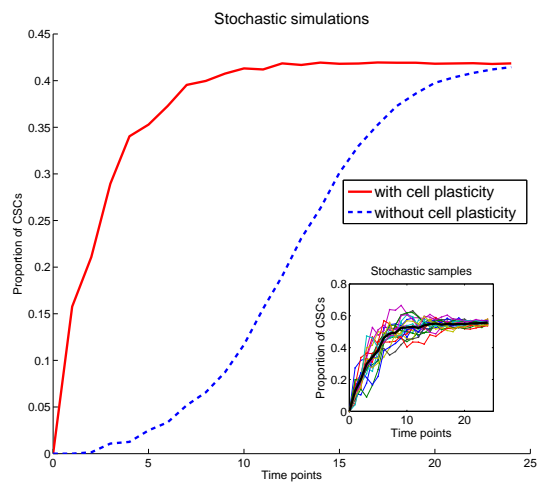}
\caption{Stochastic simulation of CSC-NSCC model: It is shown that starting from very small CSCs proportion,
the stochastic model without cell plasticity shows sigmoidal pattern growth, while the model
with cell plasticity predicts the transient increase of CSCs proportion. This is in accord with the prediction by the Eq. (\ref{CSC-NSCC}) in Fig. 2A. The small window shows the strategy of our stochastic simulation. For given parameters,
20 stochastic samples (colorful thin lines) were produced by Monte carlo simulation.
The thick black line is the representative trajectory by averaging these 20 stochastic samples.}
\end{center}
\end{figure}

\end{document}